\documentclass{eptcs}
\providecommand{\event}{ACL2 2011} 
\usepackage{acl2}
\usepackage{amsmath}
\usepackage{graphicx}
\usepackage{alltt}
\usepackage{multirow}

\newcommand{\ie}{\emph{i.e.}, }
\newcommand{\eg}{\emph{e.g.}, }
\newcommand{\e}[1]{\emph{#1}}
\newcommand{\p}[1]{\begin{normalfont}\frenchspacing\texttt{#1}\end{normalfont}}

\newcommand{\Comment}[1]{}

\title{Integrating Testing and Interactive Theorem Proving}

\author{
    Harsh Raju Chamarthi \qquad\qquad Peter C. Dillinger
    \institute{College of Computer and Information Science,\\
               Northeastern University}
    \email{harshrc@ccs.neu.edu \qquad\qquad pcd@ccs.neu.edu}
    \and
    Matt Kaufmann
    \institute{Dept. of Computer Science,\\
               University of Texas at Austin}
    \email{kaufmann@cs.utexas.edu}
    \and
    Panagiotis Manolios
    \institute{College of Computer and Information Science,\\
               Northeastern University}
    \email{pete@ccs.neu.edu}
    }

\def\titlerunning{Integrating Testing and Interactive Theorem Proving}
\def\authorrunning{H.R. Chamarthi, P.C. Dillinger, M. Kaufmann \& P. Manolios}

\begin{document}
\maketitle



\begin{abstract}
  Using an interactive theorem prover to reason about programs
  involves a sequence of interactions where the user challenges
  the theorem prover with conjectures. Invariably, many of the
  conjectures posed are in fact false, and users often spend
  considerable effort examining the theorem prover's output
  before realizing this. We present a synergistic integration of
  testing with theorem proving, implemented in the ACL2 Sedan
  (ACL2s), for automatically generating concrete counterexamples.
  Our method uses the full power of the theorem prover
  and associated libraries to
  simplify conjectures; this simplification can transform
  conjectures for which finding counterexamples is hard into
  conjectures where finding counterexamples is trivial. In fact,
  our approach even leads to better theorem proving, \eg if
  testing shows that a generalization step leads to a false
  conjecture, we force the theorem prover to backtrack, allowing
  it to pursue more fruitful options that may yield a proof. The
  focus of the paper is on the engineering of a synergistic
  integration of testing with interactive theorem proving; this
  includes extending ACL2 with new functionality that we expect
  to be of general interest.  We also discuss our experience in
  using ACL2s to teach freshman students how to reason about
  their programs.
\end{abstract}


\section{Introduction} \label{intro}

Users of interactive theorem provers such as ACL2 spend most of
their time and effort challenging the theorem prover to find
proofs of conjectures.  They may start with a high-level theorem,
only to find that a very long sequence of other
theorems must be proven before the theorem prover is convinced
that the original conjecture is in fact a theorem.

During this process users invariably challenge the theorem prover
with conjectures that are false. For example, an intermediate
lemma may be missing some non-obvious hypotheses. In such cases,
users routinely have a difficult time determining whether the
theorem prover failed because the conjecture is not true or
because the theorem prover cannot find a proof without further
user assistance.

Lightweight methods for quickly and automatically identifying
false conjectures have the potential to dramatically simplify the
interactions between users and theorem provers.

In this paper we explore the obvious, well-studied idea of using
random testing to try to find concrete counterexamples to
conjectures. A naive approach to random testing is unlikely to
find counterexamples in all but the simplest of cases. One reason
is that it is highly unlikely that random assignments will
satisfy even relatively simple hypotheses.  This is especially
true in a theorem prover for an untyped logic, like ACL2, where
all variables can take on any value.

We use a general data definition framework that is integrated
with our testing framework. Together, they enable us to
infer type information automatically from hypotheses.

Unfortunately, hypotheses are often much more complex than a
sequence of type restrictions.  Our first contribution shows how
to overcome this problem by using the full power of ACL2 to
simplify conjectures for better testing. While previous work has
suggested that subgoals generated during the proof process can be
tested independently, as far as we know, no one has ever
described or designed a system that does this automatically for a
fully featured interactive theorem prover.  The effectiveness of our
approach is magnified by the use of libraries, including not only
general-purpose libraries developed by the user community over many
years (\eg for reasoning about arithmetic and lists), but also
domain-specific libraries developed for specific applications.
Such libraries may contain rules, typically conditional rewrite rules,
as well as other strategic guidance in the form of verified
metatheoretic simplifiers, dynamically computed hints, and so-called
\emph{clause processors} to connect to external tools.

Some of the challenges we faced in integrating testing and
theorem proving are due to the powerful proof procedures in ACL2,
which can generate subgoals that differ radically from the goals
they receive. For example, proof procedures may remove and
introduce variables. How does one then take subgoal
counterexamples and turn them into counterexamples for the
original conjecture? Should we test every subgoal, or is there
some disciplined way of testing select subgoals? Proof procedures
may strengthen subgoals, making it possible to find
counterexamples to subgoals derived from \emph{true}
conjectures. They may generate implied facts that are not part of
a subgoal, but that we can profitably mine for useful
information. 

Proof procedures may also remove or modify ``type''
hypotheses (after all ACL2 is untyped, so ``type'' hypotheses are
just regular hypotheses and, therefore, subject to rewriting and
other forms of simplification).  In fact, a variable may wind up
having multiple type hypotheses associated with it; so what is
the best strategy for generating random test cases that satisfy
these hypotheses?  The way we deal with these issues is described
in Section~\ref{improved-random-testing}.

Our second contribution is to show, perhaps surprisingly, that
not only can theorem proving lead to better testing, but testing
can lead to better theorem proving. For example, suppose that the
theorem prover generalizes a goal, but subsequent testing shows
that the resulting conjecture is not valid. Then we can force the
theorem prover to backtrack so that instead of painting itself
into a corner, the theorem prover can pursue more fruitful
options that may yield a proof.  In fact, this idea can be applied
to several proof procedures within ACL2, and can also be used to
help the theorem prover choose from a set of applicable proof
steps.  We describe this idea in
Section~\ref{improved-theorem-proving}, where we also present an
example from the ACL2 regression suite.

Our third contribution, described in
Section~\ref{implementation}, consists of enhancements related to the
\emph{computed hint} mechanism of ACL2 required for the
integration of testing with theorem proving.  The first
enhancement involves changing ACL2 so that it records the reasons
for eliding variables. We need this information to generate
counterexamples for top-level conjectures from subgoal
counterexamples. This enhancement is discussed in
Section~\ref{improved-random-testing}. The remaining three
enhancements are to the \emph{computed hint} mechanism in ACL2.
Computed hints are a very powerful mechanism that allow users to
compute hints dynamically, by examining the subgoals ACL2
generates during the theorem proving process.  The first
enhancement to computed hints is that they are now given access
to various sources of derived facts that are not part of a
subgoal, but that can be quite useful for testing.  The second
enhancement, which we call {\em override-hints}, provides a kind
of meta-programming capability for computed hints that allows us
to add testing hints dynamically to interesting subgoals, without
interfering with user-provided hints.  The third enhancement to
computed hints, \emph{backtrack hints}, permits a limited form of
backtracking.
We expect that these enhancements will be of use
to the wider ACL2 community, and may be of interest to developers
of other theorem proving systems.

Our fourth contribution involves the implementation and
evaluation of the work presented in this paper. 
All of our work has been implemented in ACL2s, the ACL2
Sedan~\cite{acl2s}. ACL2s uses ACL2 as its core reasoning engine,
but was designed with particular emphasis on usability by a wide
range of users.  In particular, ACL2s provides a modern
integrated development environment in Eclipse, supports several
{\em modes} of interaction, and incorporates a powerful automated
termination analysis engine~\cite{ManoliosVroon06}. ACL2s is
freely available, and well-supported. These enhancements have
made it possible for us to use ACL2s to teach hundreds of
freshman students at Northeastern University how to reason
formally about programs.

The work in this paper is motivated by our experience teaching
college freshmen.  Even advanced freshmen have not been exposed
to the idea of program verification. However, all of the students
do know how to program and how to evaluate a program on concrete
inputs. Therefore, it is easy to explain how to falsify a
conjecture with testing: find inputs such that evaluating the
conjecture with these inputs yields {\em false}. A conjecture is
true if no such inputs exist. This is a good way of teaching
students about specification in a way that directly connects what
they know, namely evaluation, to the new notions of specification
and verification. When they first start, they often make silly
mistakes, specifying conjectures that they mistakenly think are
true. Therefore, tools that automatically falsify conjectures and
provide witnesses that students can evaluate can serve an
important pedagogical role.

We have designed our testing framework with both
beginners and experts in mind.  The interfaces are as simple as
possible. In fact, no special incantations are required to
use testing. In ACL2s, it will just happen automatically.
We have successfully used ACL2s augmented with testing in our
freshman classes.  We expect our work to make ACL2s a more useful
tool for students as well as the wider community. We briefly
discuss our experiences in Section~\ref{exp}.


\section{Related Work} \label{related-work} 
\subsection{Counterexample Generation in Interactive Theorem Provers}
Random Testing is a well-studied, scalable, lightweight technique
for finding counterexamples to executable formulas.  Random
testing has been widely adopted in the functional language
community, as seen by the recent success of QuickCheck~\cite{ClaessenH00}.
The theorem-proving community
has also embraced random testing, for example in
Isabelle/HOL~\cite{BN04}, Agda~\cite{DHT03} and
PVS~\cite{Owre06}.  
The other standard technique for generating
counterexamples for a conjecture is to use a SAT or SMT
solver. This requires translating from a rich, expressive logic
to a restricted logic with limited expressiveness. The major
constraint on such approaches is that a counterexample to the
translated formula should also be a counterexample to the
original formula. However, the absence of a counterexample does
not imply that the conjecture is true.  Some tools making use of
the above technique are Pythia~\cite{SK07}, SAT
Checking~\cite{Sumners02}, Refute~\cite{Weber08} and
Nitpick~\cite{nitpick10}. Another line of work translates to SAT
or other decidable fragments of first order logic for which
efficient decision procedures exist, but only when the original
conjecture is in fact expressible in the decidable
fragment~\cite{MS05,MS06,HuntReeber06}. ACL2 has included a
related capability since 1995, when BDDs with rewriting were integrated into
ACL2~\cite{acl2-bdd}.  The work mentioned above has the same goal
as our work: to exhibit counterexamples to false conjectures.
However, unlike our work, none of the above mentioned approaches
is a fully automated method that uses an interactive theorem
prover to generate counterexamples for arbitrary executable conjectures.

\subsection {Combining testing and theorem proving} 
One of the first convincing examples of combining testing and
proving was carried out using Agda~\cite{DHT03}, although the 
ideas for combining formal specifications (and tools) and
testing date back to at least 1981~\cite{corky81}. In Agda, 
random testing is used to check for counterexamples
to the top-level goal, and the authors suggest that
the user could also manually apply random testing to subgoals.
However, in contrast to our work, the integration of testing is not automatic.

There has been a lot of work on employing formal methods technology to
perform model-based testing since the seminal work of Dick and 
Faivre~\cite{dick1993automating}. 
We restrict our attention to
some recent work leveraging theorem provers towards this goal.
In the tool \textsc{hol-TestGen}~\cite{BW04}, specifications
are analyzed symbolically (unfolding definitions) using Isabelle/HOL 
to derive formulas in conjunctive normal form. 
To handle recursion, a \e{uniformity hypothesis}~\cite{gaudel1995testing} 
is used to bound the number of unfoldings. After minimization, the resulting 
formulas, called symbolic test-cases, are grounded using 
random or user-specified test-data generators.
Particular emphasis is put on interactive control of \e{test hypotheses}
(derived from \e{uniformity and regularity hypotheses}) for tractable testing. 
A similar tool with less focus on user interaction
is FocalTest~\cite{carlier2008functional}. FocalTest
transforms a top-level property into a set of 
elementary properties (normal form) which are independently tested by use of
random test-data generation. Transforming a specification into a
normal form can be viewed as a form of case-analysis. ACL2's 
proof procedures accomplish much more than just case-analysis;
in particular, rewrite rules programmed in ACL2 by user-specified 
lemmas or lemmas in standard libraries contribute greatly to
the effectiveness of our integrated testing.

An obvious difference between our implementation and the 
aforementioned systems is that they do not
use the result of the testing process to influence the theorem proving process.
In the context of ACL2, there has been previous work with the goal of preventing
the prover from performing ``bad'' generalizations.
In~\cite{Erickson07backtrack}, Erickson describes an extension to ACL2
allowing it to backtrack from a failed proof to alternative proof strategies.
Erickson's procedure works by initially calling the simplifier (\p{bash})
to process the original goal, returning an equivalent list of clauses. 
The procedure then attempts to refute
each clause. This involves unwinding the recursive functions to
some finite depth and sending the resulting formula to \p{bash}. 
The failure of the simplifier to prove the formulas is
used to control backtracking during generalization. Although we share
the common goal of improved theorem proving, there are a number of differences.
Adopting Erickson's approach requires significant changes to the 
main prover loop. His implementation can handle arbitrary ACL2 conjectures,
even \e{constrained} functions, unlike us. 
We employ a sound technique of using evaluation (testing) to 
control the backtracking, whereas Erickson's procedure uses the \p{bash}
simplifier, which potentially can fail to prove a \e{true} clause, thus
resulting in preventing a potentially ``good'' generalization.

Our approach seamlessly integrates ACL2 theorem proving with testing,
with a high level of automation. In the context of 
interactive theorem proving, we know of no previous
work for which reasoning and testing are tightly integrated with
each automatically informing the other.

\section{Test Generation} 
\label{datadef}
ACL2 formulas tend to be executable; hence testing in ACL2 simply
involves executing a formula under an instantiation of its free
variables.\footnote{Many other theorem provers also provide
  various levels of support for executing formulas.}  

For testing to be effective, the variables should be bound to
values satisfying the ``type-like'' hypotheses of the formula.
While ACL2 is syntactically {\em untyped}, the ACL2 value
universe is divided into 14 pairwise-disjoint ``primitive types''
which include $\{0\}$, the positive integers, the positive
non-integer rationals, the negative integers, the negative
non-integer rationals, the complex rationals, $\{nil\}$, $\{t\}$,
other symbols, null-terminated non-empty lists, conses that are
not null-terminated lists, strings, characters, and everything else (the
universe is not closed).  ACL2 users provide the prover with type
information by specifying type constraints (hypotheses such as
\p{(stringp x)}).

One cannot create new {\em types} in ACL2, in the sense that one
cannot define a new non-empty set of values that provably extends
the ACL2 value universe.  Rather, one defines a ``type'' by
defining a predicate that recognizes a subset of the ACL2
universe (\eg \p{true-listp}).

ACL2s includes a data definition framework~\cite{defdataCD} that supports and
automates the generation of such user-defined data types.  For
example, the following form defines a list of integers (\p{loi}).

\small
\begin{verbatim}
(defdata loi (listof integer))
\end{verbatim}
\normalsize

\noindent Given the above form, ACL2s will automatically generate a
\emph{type predicate} \p{loip}, which recognizes lists of
integers.  The data definition framework also supports testing by
generating \p{nth-loi}, a type enumerator that maps natural
numbers into lists of integers. In general, a \emph{type
  enumerator} for type $T$ is a surjective function from natural
numbers to $T$.  ACL2s will automatically generate a type
enumerator for any new data types defined using the data
definition framework. The type predicate and enumerator are 
 generated using a syntax-directed translation. We say that
\e{foo} is a `type', recognized by the data definition framework, 
if there exists a predicate function, \p{foop}, and
an enumerator function, \p{nth-foo}. If \e{foo} is a `type',
it can be used in a \p{defdata} form to define another `type'.
In the above example, \e{integer} is used to define a list of 
integers (\e{loi}).
 
ACL2s also proves certain theorems and
updates a global table maintaining metadata for existing
datatypes, \eg it will prove and then record that \p{loi} is a
subtype of \p{true-list}.  Below we show what \p{loip} and
\p{nth-loi} evaluate to on simple examples.

\small
\begin{verbatim}
(loip '(-1 -23 -42 7 13)) = T

(nth-loi 26945) = (24 -5 1 0)
\end{verbatim}
\normalsize

The data definition framework in ACL2s provides type enumerators
for the primitive types and for basic ACL2 types that are combinations of
primitive types (\eg the natural numbers, integers, rationals,
and lists). Each data object in the ACL2 universe is treated as a
singleton ``type'', \ie a set with just one element, the data object
itself. The type which represents all of the ACL2 universe is 
called {\em all}; every type is thus a subset of {\em all}.
Also fully supported are user-defined union
types, product types, list types, set types, record types, and enumerated
types.  ACL2s even  supports custom types (\eg prime numbers),
but then the burden of generating the enumerator falls on the
user. Table \ref{datadeft} gives a synopsis of the main features of
the data definition framework.

\begin{figure*}[ttt!]
\label{datadeft}
\begin{tabular}{|l|p{11cm}|}
\hline
\textbf{Feature} & \textbf{Example} \\ 
\hline
{\footnotesize Enumerated Types} & 
\begin{minipage}[t]{\linewidth}
{\footnotesize
\begin{verbatim}
(defdata RGB (enum '(red green blue)))

\end{verbatim}
}
\end{minipage} 
\\ \hline
{\footnotesize Union Types} & 
\begin{minipage}[t]{\linewidth}
{\footnotesize
\begin{verbatim}
(defdata BorC (oneof boolean character))

\end{verbatim} 
}
\end{minipage} 
\\ \hline
{\footnotesize Product Types} & 
\begin{minipage}[t]{\linewidth}
{\footnotesize
\begin{verbatim}
(defdata NP (cons nat (cons pos (cons neg nil))))

\end{verbatim} 
}
\end{minipage} 
\\ \hline
{\footnotesize Types with Macros} & 
\begin{minipage}[t]{\linewidth}
{\footnotesize
\begin{verbatim}
(defdata NP (list nat pos neg))

\end{verbatim} 
}
\end{minipage} 
\\ \hline
{\footnotesize List Types} & 
\begin{minipage}[t]{\linewidth}
{\footnotesize
\begin{verbatim}
(defdata loi (listof integer))

\end{verbatim} 
}
\end{minipage} 
\\ \hline
{\footnotesize Set Types} & 
\begin{minipage}[t]{\linewidth}
{\footnotesize
\begin{verbatim}
(defdata points (set (list x-pos y-pos)))

\end{verbatim} 
}
\end{minipage} 
\\ \hline
{\footnotesize Record Types} & 
\begin{minipage}[t]{\linewidth}
{\footnotesize
\begin{verbatim}
(defdata pg-tbl-entry (record (valid . boolean)
                              (protection . boolean)
                              (ppage-addr . p-addr)))

\end{verbatim} 
}
\end{minipage} 
\\ \hline
{\footnotesize Recursive Types} & 
\begin{minipage}[t]{\linewidth}
{\footnotesize
\begin{verbatim}
(defdata tree (oneof 'Leaf
                     (Node (id . symbol)
                           (left  . tree)
                           (right . tree))))

\end{verbatim} 
}
\end{minipage} 
\\ \hline
{\footnotesize Mutually Recursive Types} & 
\begin{minipage}[t]{\linewidth}
{\footnotesize
\begin{verbatim}
(defdata  
  (sexp (oneof symbol integer slist))
  (slist (oneof nil (cons sexp slist))))

\end{verbatim} 
}
\end{minipage} 
\\ \hline
{\footnotesize Custom Types} & 
\begin{minipage}[t]{0.5\linewidth}
{\footnotesize
\begin{verbatim}
(defun primep (x)
  (and (natp x)
       (= (num-divisors x) 2)))

\end{verbatim}
}
\end{minipage}
\begin{minipage}[t]{0.5\linewidth}
{\footnotesize
\begin{verbatim}
(defun nth-prime (n)
  (nth n (sieve (upper-bound n)))

\end{verbatim}
}
\end{minipage}
\\ \hline 
\end{tabular} 
\caption[]{Data definition framework - Examples\footnotemark}
\end{figure*}

To enable effective testing, one should use type predicates understood
by the data definition framework (\eg \p{loip}) to specify the types
of the free variables in the hypotheses of a conjecture.  The
corresponding type enumerator (\eg \p{nth-loi}) can then be used
to generate test samples. There are many ways in which test
generation can proceed. For example, we can enumerate test
instances up to a certain size (bounded exhaustive testing); 
we can randomly sample (random testing); we can do
both. Currently, the default is to randomly sample. 
Furthermore, the separation
of concerns between enumerators and random number generators also
gives us the flexibility to choose any kind of random
distribution. Currently we have support for pseudo-geometric and
pseudo-uniform random distributions.\footnotetext{The enumerator given
as an example for custom type \e{prime} uses the sieve of Eratosthenes, 
to find all prime numbers less than a given bound. 
One naive upper bound of \p{(nth-prime n)} is $2^{n+1}$, for $n \geq 1$.}

To show the testing framework in action, we pick a classic
example. After defining \p{rev}, the user tests that taking the
reverse of a reversed list gives back the original list:

\footnotesize
\begin{verbatim} 
(defun rev (x)
  (if (endp x)
      nil
    (append (rev (cdr x)) (list (car x)))))

(top-level-test? (equal (rev (rev x)) x))
\end{verbatim}
\normalsize
The result is the following output:
\footnotesize
\begin{verbatim}
Random testing with type alist ((X . ALL))

We falsified the conjecture. Here are counterexamples:
 -- (X 0)
 -- (X "ba")

Cases in which the conjecture is true include:
 -- (X NIL)
 -- (X (U |h|))
... 
\end{verbatim}
\normalsize

The above output snippet tells us that $x$ was randomly instantiated
with type \e{all} and for $x = 0$ and $x = $\verb|"ba"|,
the value of \p{(rev (rev x))} is not equal to the value of \p{x}.
This illustrates a common mistake made by new users: ACL2 is a
logic of total functions, but new users often assume that a
conjecture is restricted to the domain of interest when
specifying conjectures.  The logic needs all assumptions to be
given explicitly. Notice that we not only generate
counterexamples, but we also generate \emph{witnesses}: examples
which satisfy both the hypotheses (none in this case) and
conclusions of conjectures. In the above example, the witnesses are
the empty list, \p{nil}, and the symbol list, \p{(U |h|)}. 
By comparing witness and counterexamples, the user can easily add the missing
type hypothesis:

\footnotesize
\begin{verbatim} 
(top-level-test? (implies (true-listp x)
                          (equal (rev (rev x)) x)))
\end{verbatim}
\normalsize
{\em Top-level-test?} now reports only witnesses:
\footnotesize
\begin{verbatim} 
Random testing with type alist ((X . TRUE-LIST))

Cases in which the conjecture is true include:
 -- (X (23 -1 0))
 -- (X (|a| 0 NIL))

We tried 100 random trials, 100 (99 unique) of which satisfied the hypotheses. 
Of these, none were counterexamples and 99 were witnesses.
\end{verbatim}
\normalsize

Notice that in the original conjecture no type restriction is
specified. Hence, random instances for {\tt x} are selected from
the entire ACL2 universe, \ie {\tt x} is of type {\em all}.  In
the modified conjecture, however, the framework automatically
extracts the type restriction that {\tt x} is a {\em true-list}
from the hypothesis and generates only examples of the desired
type. Thus, we are guaranteed that no test passes trivially,
merely because the hypotheses were not satisfied; \ie there are
no vacuous witnesses. The framework ``understands'' and
syntactically extracts two types of type restrictions:
\begin{enumerate}
\item \emph{Datatype hypotheses} such as \p{(loip x)} where \emph{loi}
is a `type', \ie it has a corresponding enumerator 
function (\ie \p{nth-loi}).
\item \emph{Equality hypotheses} such as \p{(equal x 42)} which is the
strongest type restriction possible, where a variable can take only one 
value (the case of a singleton type).
\end{enumerate}
Often type restrictions are more complex than \e{datatype
  hypotheses}; we consider a variation of another 
classic example~\cite{dick1993automating} below. A
triangle is a triple of positive integers (recognized by type
predicate \p{posp}), representing its three sides, with each side
less than the sum of the other two sides.

\footnotesize
\begin{verbatim}
(defdata triple (list pos pos pos))

(defun trianglep (v)
  (and (triplep v)
       (< (third v) (+ (first v) (second v)))
       (< (first v) (+ (second v) (third v)))
       (< (second v) (+ (first v) (third v)))))
\end{verbatim}
\normalsize

The \p{shape} function determines whether its argument is an equilateral,
isosceles, scalene or illegal triangle.

\footnotesize
\begin{verbatim}
(defun shape (v)
  (if (trianglep v)
      (cond ((equal (first v) (second v))
             (if (equal (second v) (third v))
                 "equilateral"
               "isosceles"))
            ((equal (second v) (third v)) "isosceles")
            ((equal (first v) (third v)) "isosceles")
            (t "scalene"))
    "error"))
\end{verbatim}
\normalsize

Consider the conjecture that there are no isosceles triangles whose third
side is the product of the other two sides and is greater than 256.

\footnotesize
\begin{alltt}
(top-level-test? 
 (implies (and (triplep x)
               (trianglep x)
               (> (third x) 256)
               (= (third x)
                  (* (second x) (first x))))
          (not (equal "isosceles" (shape x)))))

\ensuremath{\Longrightarrow}

Random testing with type alist ((X . TRIPLE))

We tried 10000 random trials, none of which satisfied the hypotheses.
\end{alltt}
\normalsize

\noindent Straightforward random testing (\p{top-level-test?})
fails miserably, because even though we pick up that \p{x} is of
type \e{triple} and randomly instantiate it, it is very hard to
satisfy the extra constraints on \p{x}. Consider the probability
of finding a counterexample to the conjecture by randomly
generating positive integer values $a$, $b$, and $c$ for the first, second, and
third side of the triangle, respectively. Let us assume
that we are using a uniform distribution over the numbers $[1..k]$.
For the case that $c > 1$ is equal to one of the other two
($a$ and $b$), the probability that we guess a
counterexample (ignoring the condition $c > 256$)
is $\frac{2}{k^2}$ because one of
$a, b$ has to be equal to $c$ and the other has to be equal to
1. For the case that $a=b$, then $c$ must be a square number and then there is
only one choice for $a, b$, namely $\sqrt c$, so the
probability that we select a counterexample is $\leq
\frac{1}{k^2}$. Once we take the $c > 256$ constraint into
account, we see that the probability of generating a
counterexample is less than $\frac{1}{32,678}$.  A constraint
solver might help find counterexamples, but this is an
undecidable problem in general. In the next section, we show how
to make use of an interactive theorem prover to tackle complex
constraints like the above. 

\textbf{Enable testing in ACL2}: 
To use the testing framework in its minimalist setting, 
submit, in an ACL2 session, the following two 
forms\footnote{Typically, if you install ACL2s (eclipse plugin),
 \e{acl2s-modes-src-dir} is located in \p{<eclipse-dir>/plugins/}.
}.

\small
\begin{verbatim}
(include-book "<acl2s-modes-src-dir>/acl2-datadef/top")
(set-acl2s-random-testing-enabled t)
\end{verbatim}
\normalsize

\section{Improved Random Testing with Theorem Proving} \label{improved-random-testing}
In this section, we show how to use the full power of the ACL2
theorem prover to simplify conjectures for better testing.  The
main idea is to let ACL2 use all of the proof techniques at its
disposal to simplify conjectures into subgoals, and then to test
those subgoals. The challenge is that ACL2 employs
proof procedures that often generate radically transformed
subgoals. We describe some of these issues and our solutions in
this section.


First, let us step back and quickly review the organization of
the ACL2 theorem prover.  ACL2 keeps track of a set of goals
to be proved, starting with the top-level
conjecture. A goal may be processed by a collection of proof
techniques, each of which uses the \emph{world}, which is a database
containing all the current axioms, theorems, and definitions. These
proof techniques are tried on a given goal,
in order, until one succeeds by producing a (possibly empty)
set of new subgoals.  If none succeeds, then the goal is pushed into a {\em pool} of
formulas to be proved by induction. This organization is called the
\emph{waterfall}~\cite{acl,car,waterfall}. The original conjecture is
proved when the pool has been fully drained.

Space limitations do not allow us to describe fully the
waterfall and its proof techniques. Instead we
will focus on two of the proof techniques in this section, 
\e{simplification} and \e{destructor elimination}. 
Simplification is quite complicated. It includes
decision procedures for propositional logic, equality,
uninterpreted functions, and rational linear arithmetic.  It uses
type information and forward chaining rules to deduce a
\emph{context} of derived facts. It uses conditional rewriting
rules and metafunctions (which can be thought of as
user-provided, verified theorem provers). It uses
if-normalization to convert formulas to a set of equivalent (but
simpler) formulas.
Such simplification techniques are carefully controlled using
heuristics developed over many years.

The first issue is what subgoals to test. We do not test every
subgoal, because simplification may generate subgoals that can be
further simplified. Instead, we test
\emph{checkpoints}, subgoals that users of ACL2 are encouraged to
examine when their proof fails~\cite{car}. Only subgoals that
cannot be further simplified are identified as checkpoints, and
ACL2 has a mechanism
that supports examination of key checkpoints. Testing at
checkpoints makes sense because case analysis has been applied,
providing more specific guidance in locating counterexamples.

The next issue is that ACL2 proof procedures may remove and/or
introduce variables. For example, simplification can decide to
replace a variable by an equivalent expression.  The
\emph{destructor-elimination} proof technique may remove and
introduce variables at the same time, as happens when ACL2 tries
to prove the following conjecture, which is identical to the one
from the previous section.

\footnotesize
\begin{verbatim}
(thm (implies (and (trianglep x)
                   (> (third x) 256)
                   (= (third x)
                      (* (second x) (first x))))
              (not (equal "isosceles" (shape x)))))
\end{verbatim}
\normalsize

\noindent We note that our integration of testing into ACL2s does not
require new syntax to be learned; users invoke \e{thm} as
before.  The prover opens up the definitions of \p{shape},
\p{trianglep} and \p{triplep} and uses case analysis, reducing
the above ``Goal'' to three subgoals.  After several
simplification steps and a few rounds of destructor elimination
on one of these subgoals,
we have the following subgoal, for which testing easily yields a counterexample.
\Comment{
 The following subgoal is
the result of simplifying the third subgoal derived from the case
\p{(equal (first v) (third v))} in the definition of \p{shape}:

\begin{small}
\begin{verbatim}
Subgoal 3''
(IMPLIES (AND (CONSP X)
              (INTEGERP (CAR X))
              (< 0 (CAR X))
              (CONSP (CDR X))
              (CONSP (CDDR X))
              (NOT (CDDDR X))
              (< 1 (* 2 (CAR X)))
              (< 256 (CAR X))
              (EQUAL 1 (CADR X))
              (NOT (EQUAL (CAR X) 1)))
         (NOT (EQUAL (CAR X) (CADDR X)))).
\end{verbatim}
\end{small}

Note: \p{first, second and third} are just aliases for the list
selector functions \p{car, cadr, caddr} respectively.  This is a
crucial simplification, and is a result of applying various
rewrite rules from the arithmetic libraries.  Notice that there
are no \e{datatype hypotheses} or \e{equality hypotheses} we can
extract at this point. Since {\tt X} represents a value built
with three cons operations, ACL2 performs three rounds of
destructor elimination, where finally {\tt X} is thrown away and
four new variables are introduced.  The replacements of terms
{\tt (CAR X)}, {\tt (CADR X)}, {\tt (CADDR X)}, and {\tt (CDDDR
  X)} by the variables {\tt X1}, {\tt X3}, {\tt X5}, and {\tt
  X6}, respectively, allows our testing framework to extract type
constraints. This step further helps the theorem prover to
simplify the subgoal, using evaluation and primitive type
reasoning; and testing then yields a counterexample.
}

\footnotesize
\begin{verbatim}
Subgoal 3'4'

(IMPLIES (AND (INTEGERP X1)
              (< 0 X1)
              (< 1 (* 2 X1))
              (< 256 X1))
         (EQUAL X1 1)).

Random testing "Subgoal 3'4'" with type alist ((X1 . POS))

We falsified the conjecture. Here are counterexamples:
 -- (X (429 1 429))
...
\end{verbatim}
\normalsize

\noindent
Notice that the theorem prover simplified away two variables representing
two sides of the triangle, thus drastically simplifying the constraints.
As we saw previously, for the original conjecture the probability of
finding a counterexample if we randomly assign positive integers
to the three sides using a uniform distribution over $[1..k]$ was
$\leq \frac{2}{k^2}$. By using the theorem prover, we generated
the subgoal above, where the probability of finding a
counterexample approaches 1 as $k$ goes to $\infty$.
Apart from case-analysis and destructor elimination, 
the primary simplification is due to
the presence of libraries of lemmas (notably arithmetic-5~\cite{KrugLib}).
  In this respect
interactive theorem proving has a huge advantage over other tools
routinely combined with testing, especially considering the fact
that most interactive theorem provers have good library support.

Experience suggests that presenting a counterexample such as
\p{X1 = 429}, which falsifies a subgoal but contains a variable
({\tt X1}) occurring in the subgoal but not in the original goal,
is less useful than a counterexample to the original goal.  In
order to construct counterexamples for the original conjecture
from counterexamples for subgoals, we automate maintenance of a
\emph{testing-history} data structure in ACL2.  This structure
associates each goal with its parent and maintains a mapping from
variables appearing in its parent to expressions over the
variables appearing in the child subgoal. For example, after
destructor elimination on the above example, we would record that
\p{X} maps to \p{(CONS X1 X2)}.  \Comment{, \p{X2} maps to
  \p{(CONS X3 X4)} and so on, and after the final simplification,
  we would record that \p{X3} maps to \p{1} and \p{X5} maps to
  \p{X1}. } Sometimes a variable is completely elided away
(consider the hypothesis \p{(EQUAL X X)}), in which case we
arbitrarily assign it the symbol \p{?}, denoting a
\e{don't-care}.  This information allows us to propagate child
goal counterexamples upward, to obtain a complete counterexample
to the top-level conjecture.\footnote{For some proof
  procedures---generalization, fertilization and induction---it
  can be hard to obtain a top-level counterexample, and we may
  fail to do so.  Indeed, such a counterexample does not exist
  when generalization changes a theorem into a non-theorem.}

Another issue is that since ACL2 is untyped, it may decide to
throw away datatype information in a hypothesis, say because it
is implied in the current context. While we can sometimes recover
this information, we would like a guarantee of \e{datatype
  monotonicity}: subgoals do not wind up with less type
information their parents. To that end, we record in the
testing-history data structure another mapping from variables in
the subgoal to a list of type restrictions that the variable must
satisfy at that subgoal and its descendants. 
The reason we have a list is that we have several type
restrictions on the same variable, arising from either several
datatype/equality hypotheses in the subgoal itself, type
information from ancestor goals, or the theorem prover
itself.\footnote{ACL2 uses a type reasoning
  mechanism which can be customized by user input in the form of
  rules.} The list may grow as we move from a goal to its
subgoals. For example, consider a goal in which a variable may be an
integer or a string. After case analysis we may wind up with two
goals: in one the variable is assumed to be a string; in the
other it is assumed to be an integer.
 
If we have several type restrictions on a variable, we would like
to use all available information for the generation of random
tests. For example, it is desirable to determine automatically
the minimal datatype that the variable satisfies. For built-in
datatypes like \e{Nat} and \e{Integer}, ACL2 already can determine this
information, but for custom datatypes and for datatypes
constructed using our testing framework (\e{defdata}), there are
several complications. First, a minimal datatype need not exist,
\eg consider the case in which the variable satisfies two datatypes,
but there is no datatype corresponding to the
intersection. Second, custom datatypes make this an undecidable
question, \eg the proof that type $T_1$ corresponds to the
intersection of types $T_2$ and $T_3$ can be arbitrarily hard to establish.
In order to deal with these issues we maintain 
a \emph{defdata subtype graph}. The vertices are the 
known data definitions and if 
there is an edge between $T_1$ and $T_2$ then $T_1 \subseteq T_2$
(we are abusing notation here by using $T_i$ to denote the subset
of the ACL2 universe satisfying data definition $T_i$). This is a
directed graph. Notice that nothing stops us from having two data
definitions that have exactly the same elements.  We allow users
to add edges to this graph by proving that one type subsumes
another using {\tt defdata-subtype}, for example as follows.

\begin{small}
\begin{verbatim}
(defdata-subtype triple proper-cons)
\end{verbatim}
\end{small}

We use the graph by first computing strongly connected
components. Nodes in the same component are provably
equivalent. We then compute the transitive closure of the
resulting dag and can use this information to help select the smallest
type associated with a variable. 

\section{Improved Theorem Proving with Random Testing}
\label{improved-theorem-proving}

In this section we describe briefly a novel use of testing to direct an
automated theorem prover.  For more details about how this works, see
the discussion of {\em backtrack hints} in Section~\ref{implementation}.

The ACL2 proof engine relies on {\em proof processes} to replace a
given goal by a list of goals, such that if each goal in that list is
provable then the given goal is provable.  One such proof process is
{\em generalization}, which replaces a goal $G$ by a single new goal,
$G'$, such that $G$ is an instance of $G'$.  It is well-known in the
ACL2 community that generalization often produces non-theorems from
goals that are theorems.  Thus, one will find numerous hints in the
ACL2 regression suite, placed manually by ACL2 users, that turn off
generalization.

In particular, as an example, consider the following lemma%
\footnote{The relevant source file can be found at
{\tt
  books/\-workshops/\-2003/\-cowles\--gamboa\--van\--baalen\_\-matrix/\-support/\-matrix.lisp} in the ACL2 regression suite.}
proved as part of an effort to formalize matrix algebra in ACL2~\cite{03-gamboa-arrays}:

\footnotesize
\begin{verbatim}
(defthm m-=-row-1-implies-equal-dot-2
  (implies (and (m-=-row-1 M2 M3 n p)
                (integerp p)
                (integerp j)
                (>= j 0)
                (>= p j))
           (equal (dot M1 M2 m n j)
                  (dot M1 M3 m n j)))
  :hints   (("Goal" :do-not '(generalize) ...)))
\end{verbatim}
\normalsize

If the hint {\tt :do-not '(generalize)} is removed, then the proof
fails because a goal is generalized to one that is no longer valid.

We would like to retain the occasional win we get from generalization,
but with fewer defeats such as the one mentioned above. Testing helps 
us by triggering backtracking, as follows.  We can arrange,
as described in Section~\ref{implementation}, for testing to be
triggered after a generalization.  If testing finds a counterexample,
then the generalization is discarded, ACL2 backtracks to the
state before the generalization, and from there it  proceeds
without generalizing that particular subgoal.  In a proof attempt for the above
example, testing prevented six attempts at generalizing a valid goal
to an invalid goal by finding a counterexample in each case, and the
proof succeeded.

Note that 
random testing might
lead to unstable proofs, where bad generalizations are discarded on
some runs of the theorem prover, but not on others, depending on
whether random sampling was successful in finding a counterexample or not. 
One solution is to fix a global constant to be used as the initial
random seed for testing a \p{defthm} form. A less restrictive solution is
to switch to bounded exhaustive testing making sure the bound 
on the number of tests is fixed to some global constant.

The example above demonstrates that testing can assist the proof
activity, by allowing the gainful use of ``dangerous'' proof
techniques (like generalization) while avoiding some of their
pitfalls.

\section{ACL2 Enhancements}\label{implementation}

In this section we explain how we exploit the ACL2 {\em hints}
mechanism to generate and evaluate tests, when appropriate.  These
ACL2 enhancements were introduced with Version 3.6 (released August,
2009).

ACL2 has long had a {\em computed hints}
mechanism~\cite{computed-hints} that allows proof hints to be computed
dynamically --- that is, during a proof attempt ---
as a function of information pertaining to the
current goal.  It may thus seem that such a mechanism is well-suited
to the integration of testing and proving, using {\em testing hints}
that may direct evaluation of the current goal in various
environments.

In order to understand why computed hints were not quite sufficient for
that purpose, we must first understand the basic structure of an ACL2
proof attempt.  Recall the {\em waterfall}, discussed in
Section~\ref{improved-random-testing}.  There is at any time a current
goal, which is initially the formula submitted for proof.  This goal
is handed to a fixed sequence of {\em proof processes}, including a
{\em simplification} process and, later in the sequence, a {\em
generalization} process.  Each process can fail or succeed on the
current goal, $G$.  The first one that succeeds replaces $G$ by a list
of goals, whose provability implies the provability of $G$.  If none
of the processes applies, then the goal is ``pushed'' for later proof
by induction.

When a goal becomes the current goal, ACL2 searches through the
available hints until it finds an appropriate hint structure (if any)
to apply to that goal.  This hint structure is applied at the ``top''
of the waterfall, that is, before the first process is attempted on
the current goal --- {\em not} as each new proof process is attempted
on that goal.

There are two problems with this approach, which we describe in turn
below.  First, in order to use theorem proving for improved testing as
described in Section~\ref{improved-random-testing}, we need to apply
not only the testing hints, but we also need to apply the user's
original hints so that the intended proof is not adversely affected by
the testing hints.  Yet, as described above, at most one hint
structure is chosen.  Second, in order to use testing for improved
theorem proving as described in
Section~\ref{improved-theorem-proving}, we use testing hints to
backtrack, so we want to be able to apply hints {\em after} a goal has
been processed and created proposed child goals.  Yet, as described
above, hints are processed at the {\em top} of the waterfall, at which
time the resulting child goals have not yet been computed.

The first problem, to allow the application of a user's hint together
with testing hints, is conceivably solvable by using computed hints to
merge hints; but this would be awkward.  Instead, this work has
inspired a new, very general utility: a new hint mechanism for
specifying easily how to {\em modify} hints selected for goals.  These
{\em override-hints}~\cite{override-hints} are expressions to evaluate, each of which can
mention the variable {\tt HINT-SETTINGS}, which is bound initially to
reflect the hint structure selected for the goal (or {\tt nil} if none
is found).  Then each override-hint is evaluated in turn, where the
result of each is the value of {\tt HINT-SETTINGS} used for evaluation
of the next override-hint.  The final such result is supplied as the
hint to use for the goal.  A utility {\tt add-override-hints} allows
the user to add override-hints to the global environment.

We turn now to the second problem: how to support backtracking as used
in Section~\ref{improved-theorem-proving}, in which the ACL2 hint
mechanism can discard a harmful attempt at generalization.  Such a
capability requires knowing that generalization is the applicable
proof process, as well as knowing the goal resulting from the
generalization.  The hint mechanism applied at the top of the waterfall would be at best awkward
to use; one would have to figure out how to invoke the waterfall
explicitly to predict what will happen, and then generate a hint based
on that result.

Instead, this work has inspired the addition to ACL2 of a {\em
backtrack hint} mechanism~\cite{backtrack}.  A backtrack hint is applied {\em after} a
proof process has been applied to a goal, $G$.  The
hint's value is an expression that can refer to variables {\tt
clause-list} and {\tt processor}.  That expression is then evaluated
in an environment in which these two variables are bound respectively
to the list of resulting goals and the proof process that has just
been applied.  The evaluation result is either the special value {\tt
nil}, indicating that the backtrack hint is to be ignored, or an
object specifying the hint structure to be applied to $G$.  In the
latter case, the clause list resulting from the proof process is
discarded, and $G$ is sent back through the waterfall with the new
hint structure.

Let us see how backtrack hints support improved theorem proving as
described in Section~\ref{improved-theorem-proving}.  Consider

\footnotesize
\begin{verbatim}
(add-default-hints '((test-gen-checkpoint)))
\end{verbatim}
\normalsize

\noindent
which indicates that the arity-0 function {\tt test-gen-checkpoint} is
to be applied to a goal in order to generate a hint structure.
Evaluation of the expression {\tt (test-gen-checkpoint)}, in turn,
generates a backtrack hint.  The code below specifies that this
backtrack hint should be applied to every goal, not just the current
goal, by using {\tt :computed-hint-replacement t}.  That backtrack
hint says that if the proof process that applies to the goal is the
generalization process, then ACL2 should run our testing apparatus to
look for a counterexample.  If there is a counterexample (i.e., if
{\tt res} is true), then the backtrack hint generates a hint structure
specifying that the goal should be re-tried with generalization turned
off.

\footnotesize
\begin{verbatim}
(defun test-gen-checkpoint ()
  `(:computed-hint-replacement t
    :backtrack
    (cond
     ((eq processor 'generalize-clause)
      (er-let*
       ((res (test-clause (car clause-list) state)))
       (value (cond (res '(:do-not '(generalize)))
                    (t nil)))))
     (t (value nil)))))
\end{verbatim}
\normalsize

Finally, we remark that override-hints are useful in combination with
backtrack hints: instead of the {\tt add-default-hints} form described
above, an override-hint can be used with a new version of {\tt
test-gen-checkpoint}.  This new version {\em extends} the existing
hint structure with a suitable backtrack hint, where that backtrack
hint also extends the existing hint structure (so that if the
generalization is discarded, then the user's hints are still
respected).

\footnotesize
\begin{verbatim}
(add-override-hints
 '((test-gen-checkpoint hint-settings)))
\end{verbatim}
\normalsize

\section{Experiences}\label{exp}

We describe two experiences using the counterexample generation
capabilities of ACL2s. One involves using counterexamples to teach
students how to reason about their programs and the other
involves an example from an expert.

For several years, we have been teaching freshman students at
Northeastern University how to reason about programs. We have
used ACL2s and it has been an invaluable teaching aid. One place
where students often struggle is in writing specifications. They
sometimes make logical and conceptual mistakes, and they often
omit required hypotheses. Therefore, they often try to prove
conjectures that are false. In large part, the motivation for
this work was to help students by providing them with
counterexamples. In this, we have succeeded because our testing
framework tends to find counterexamples easily. The
counterexamples allow students to see what is wrong with their
conjectures, in terms they readily understand. Without the
counterexamples we generate, students are left trying to
determine whether they need more lemmas or whether their
conjectures are false, a skill that takes time to develop. From a
usability point of view, we note that our testing framework has
the nice property that it does not incur any cognitive load on
the user. Students do not have to enable testing; they do not
have to give it hints; they do not have to invoke it. Testing is
just an invisible, natural part of the theorem proving process.

Our counterexample generation can also be fruitfully used by
experts. Consider the following conjecture:

\footnotesize
\begin{verbatim}
(thm (implies (and (real/rationalp a)
                   (real/rationalp b)
                   (real/rationalp c)
                   (< 0 a)
                   (< 0 b)
                   (< 0 c)
                   (<= (expt a 2) (* b (+ c 1)))
                   (<= b (* 4 c)))
              (< (expt (- a 1) 2) (* b c))))
\end{verbatim}
\normalsize 

An ACL2 expert who uses ACL2 to reason about industrial designs
asked the ACL2 mailing list for help in proving (his formulation of)
the above conjecture with ACL2.
However, his conjecture was missing a hypothesis.
ACL2s immediately came up with a counterexample:

\footnotesize
\begin{verbatim}
Random testing ``Subgoal 1" with type alist 
((A . RATIONAL) (B . RATIONAL) (C . RATIONAL))

We falsified the conjecture. Here are counterexamples:
 -- (A 1/7), (B 2/11) and (C 2/9)
...
\end{verbatim}

\normalsize We indicated that the conjecture was not true, and
the expert quickly strengthened one of the hypotheses to $a
\geq 1$. Three ACL2 users then quickly replied with successful proofs,
and an ACL2s proof took advantage of automatically included libraries for
reasoning about arithmetic~\cite{KrugLib}. In fact, with a little bit
of trial and error, we were able to generalize the theorem by
removing hypotheses \p{(< 0 b)} and \p{(< 0 c)}. Using binary
search we also replaced \p{(<= 1 a)} with \p{(< 3/4 a)}. We know
that the bound is tight because \p{(<= 3/4 a)} leads to a
counterexample. We also know no further hypotheses can be removed
because doing so leads to our framework generating
counterexamples.

The moral of the story here is that if the expert had submitted his
conjecture to ACL2s,
he would have immediately been presented
with a counterexample. He would have then fixed the hypothesis
and would have been rewarded with a QED.

\section{Conclusions and Future Work} \label{conc-future-work}

Our work integrates random testing seamlessly with theorem
proving, resulting in a system with powerful, automatic testing
capabilities and more guidance in proof discovery.  We identify
four contributions. First, we show how to use the power of an
interactive theorem prover for better testing. Second, we show
how to use testing to make interactive theorem proving more
powerful and automatic. Our third contribution is a set of
enhancements to the ACL2 theorem prover's hint mechanism that
were developed to support this work and that we expect will be of
interest to the wider interactive theorem proving community. Our
fourth contribution is the implementation of the above ideas in
the ACL2 Sedan, a freely available theorem prover which has been
used to teach several hundred freshman students how to reason
about programs.

For future work, we plan to explore more powerful algorithms for
generating counterexamples.
We also plan to explore support for intersection types and more
powerful methods for determining where in the partial order of
types user defined datatypes belong, along with automatic
generation of rules to relate different such types.  (J Moore
[private communication] is exploring similar support and methods,
which may lead to collaborative solutions.)  Finally, we plan to
explore the integration of constraint solving and decision
procedures into our framework, for even better counterexample
generation.

\section{Acknowledgements}

This research was funded in part by NASA Cooperative Agreement
NNX08AE37A, DARPA Grant 10655648, and NSF grants CCF-0429924,
IIS-0417413, CCF-0438871, CNS-0910913, and CCF-0945316. Kaufmann also thanks
the Texas -- United Kingdom Collaborative for travel support to
Cambridge, England, and the Computer Laboratory at the University
of Cambridge for hosting him during preliminary preparation of
this paper.

\bibliographystyle{eptcs}
\bibliography{paper}

\end{document}